\documentclass[twocolumn]{article}

\usepackage[width=17cm,height=22cm]{geometry}
\usepackage[french]{babel}
\usepackage[utf8]{inputenc}
\usepackage{fancyvrb}
\usepackage{authblk}
\usepackage{hyperref}
\usepackage{amsfonts}
\usepackage{xcolor}
\usepackage{tikz}
\usetikzlibrary{shapes.geometric, arrows}
\usepackage{amsmath}
\usepackage{algpseudocode}
\usepackage{algorithm}
\usepackage{adjustbox}

\tikzstyle{startstop} = [rectangle, rounded corners, minimum width=2cm, minimum height=1cm,text centered, draw=black, fill=red!30]
\tikzstyle{io} = [trapezium, trapezium left angle=70, trapezium right angle=110, minimum width=2cm, minimum height=1cm, text centered, draw=black, fill=blue!30]
\tikzstyle{process} = [rectangle, minimum width=2cm, minimum height=1cm, text centered, draw=black, fill=orange!30]
\tikzstyle{decision} = [rectangle, minimum width=2cm, minimum height=1cm, text centered, draw=black, fill=green!30]
\tikzstyle{preparation} = [rectangle, minimum width=2cm, minimum height=1cm, text centered, draw=black, fill=green!30]
\tikzstyle{arrow} = [thick,->,>=stealth]  
\tikzstyle{arrow} = [thick,->,>=stealth]
\tikzstyle{dotted_arrow} = [thick,->,>=stealth, dashed]

\title{Multi-scale streaming anomalies detection for time series}
\author[1]{B Ravi Kiran\thanks{beedotkiran@gmail.com}}
\affil[1]{CRIStAL Lab, UMR 9189, Université Charles de Gaulle, Lille 3 }

\begin{document}
\maketitle

\begin{abstract}
In the class of streaming anomaly detection algorithms for univariate time series, 
the size of the sliding window over which various statistics are calculated is an important parameter. 
To address the anomalous variation in the scale of the pseudo-periodicity of time series, we define a streaming
multi-scale anomaly score with a streaming PCA over a multi-scale lag-matrix.  
We define three methods of aggregation of the multi-scale anomaly scores. We evaluate their 
performance on Yahoo! and Numenta dataset for unsupervised anomaly detection benchmark. 
To the best of authors' knowledge, this is the first time a multi-scale streaming anomaly detection has been
proposed and systematically studied. 
\end{abstract}

\medskip

\noindent\textbf{Mots-clef}: Multi-scale, Anomaly detection, Streaming, PCA

\section{Introduction}
\label{sec:lentete}

Anomalies in time series are defined as points which deviate in a significant way from 
a certain model. For instance this can be the deviation from the local mean value of the time series. 
Anomaly detection is an important problem in industrial process control and for biomedical applications.
There are few common methods used in the detection of anomalies once we define 
the representation of the time series at each point. One can either look at the 
distribution of the distances of the k-nearest neighbour of a given point ~\cite{LOFBreunig2000},
which basically measures for relative density of neighbours around a point. One can evaluate the 
errors of prediction in auto-regressive models~\cite{Laptev2015YahooBenchmark} which models
the predicted value as a linear function of the past. Finally one could reconstruct the input 
point or vector while tracking the principal subspace ~\cite{PapadimitriouSPIRIT2005}, this helps 
locate the dominant pattern in a time series. 

Streaming anomaly detection consists of detecting anomalies on the fly, where one has
access only to the current time sample and all its past values at any time. 
This is frequently applicable to large data processing requirements at each time instant, 
that storing models or representations of these signals are not feasible. 
Algorithms designed for streaming should update their parameters 
on the arrival of each new point so as to adapt to changes in the time series. 

A point anomaly is the value of the time series which deviates from the rest of 
the time series. An anomalous pattern is a window of values of 
that deviates from the rest of the windows of the time series. One can also note that
the point anomaly is also a anomaly pattern when the window is point sized.

\paragraph{Multiscale Anomalies} Given different time series we aim to take into account
the variation in the window sizes at which anomaly patterns occur in the series. 
In this paper we address the problem of designing a multiscale streaming anomaly detector.
We also study the multiscale approximation of the the moving window of the time series by the
haar wavelet transformation. The second goal is to study the effect principal subspace tracking
over the haar basis.

\begin{figure*}
  \centering
  \includegraphics[width=0.95\textwidth]{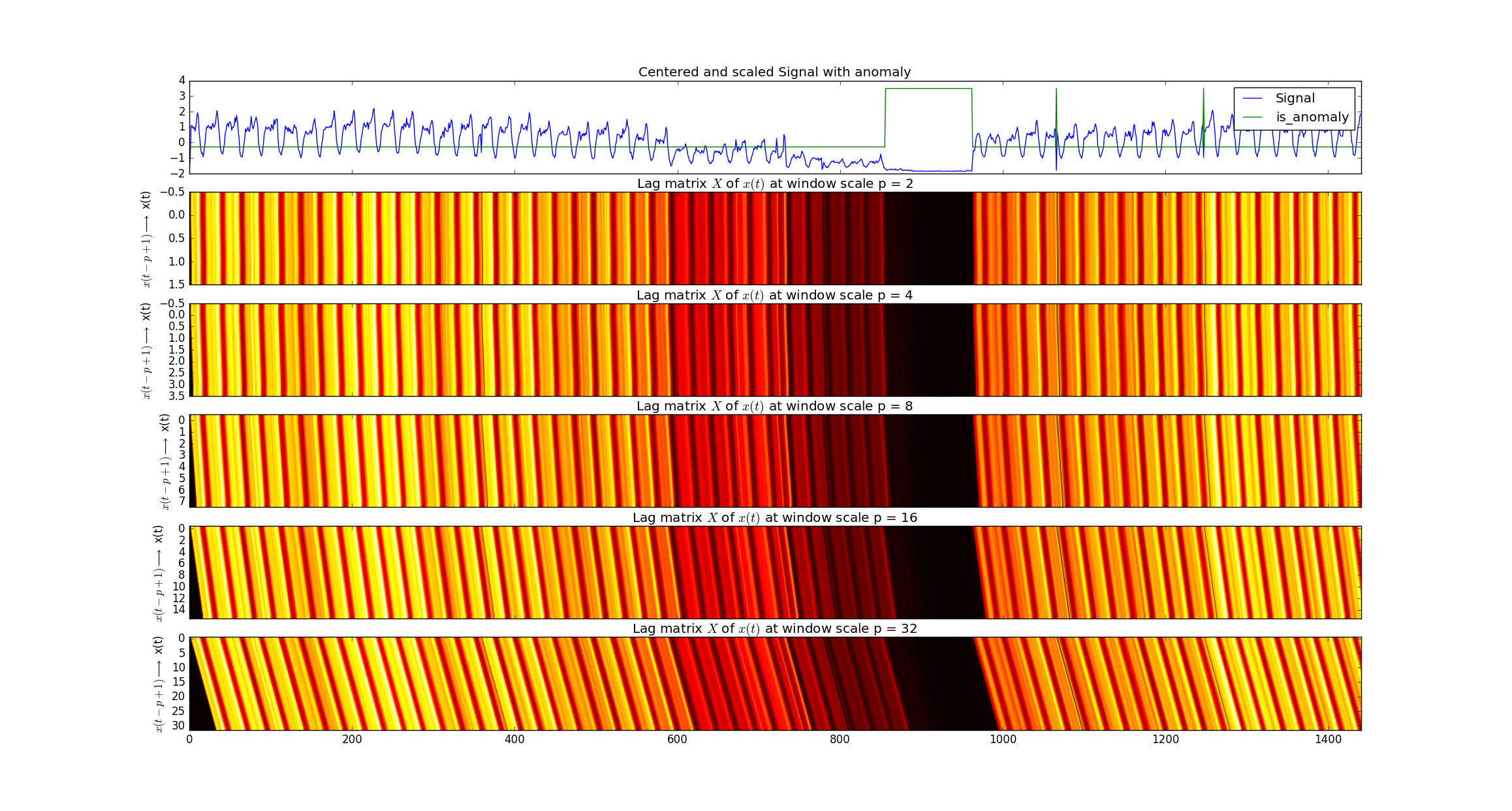}
  \caption{Example input time series and their corresponding lag-matrices $X_t^p$ for 
  different scales of moving window $p=2, 4, 6, 8, 32$. The multi-scale streaming
  PCA evaluates the principal subspace at each scale $p$. The reconstruction error
  across all scales $p$ gives the final multi-scale anomaly score $\pmb{\alpha}_t$.}
  \label{fig:lagmatrix example}
\end{figure*}

\paragraph{Previous work}
\cite{Laptev2015YahooBenchmark} proposed several linear predictive models (Autoregressive, Kalman filter) 
followed by an anomaly score filtering (by $k\sigma$ rule, or local outlier factor scores introduced 
by~\cite{LOFBreunig2000}) to detect anomalies. The authors also published a dataset to evaluate 
unsupervised anomaly detectors on real and synthetic data. This is for off-line time series data
though we shall use the annotations available to evaluate the anomaly detection performance.

The Numenta Anomaly Benchmark (NAB)~\cite{LavinA15_Numenta_NAB} provides an 
evaluation of streaming time series anomaly detection algorithms.
The algorithms include hierarchical temporal memory, Twitter Advec and Skyline, each of 
which individually are anomaly detection packages. We shall not evaluate within the 
NAB benchmark \footnote{\href{https://github.com/numenta/NAB}{https://github.com/numenta/NAB}} for the moment, 
but we are planning a detailed evaluation in the future. \cite{OnlinePCAReview2015} provides
an overview of incrementally calculate principal components, while also discussing the choice
of learning rate parameter. \cite{anomaly_princomp_tracking_2010} use a streaming subspace 
tracking algorithm that provides a rank adaptive and numerically stable streaming PCA algorithm for anomaly detection. 

There are two studies closest to our work : Firstly \cite{papadimitriou2006optimal} evaluate a multiscale
streaming PCA to extract the most ``representative'' window from an lag-embedding of
an univariate time series for different window sizes. The algorithm was used to track 
correlations at multiple scales. While \cite{mutliscale_2008} perform multiscale anomaly detection
by searching for infrequent windows in the input time series, they do not evaluate anomalies on-line.

There has also been recent work in the evaluation of on-line empirical mode decomposition
(EMD) paper \cite{Fontugne2017Icassp} which evaluates the empirical modes in streaming.
There has also been recent work on use matrix sketching to approximate
wavelet coefficients in a streaming set up \cite{cormode2006fast}.

\section{Background and Notations}
\paragraph{Notations}: The univariate time series are denoted by $x(t)$, and $T$ denotes the 
number of samples, and $p$ the lag window size. We embed the window of past $p$ 
values of the univariate time series to construct a $p$-dimensional sample $X_t^p$ 
of the multivariate time series:
\begin{equation}
    X_t^p = [x_{t}, x_{t-1}, \ldots, x_{t-p+1}]^T \in \mathbb{R}^p
\end{equation}
When $t<p$, the missing values in $X_t^p$ repeat most recent time sample. 
The vectors are stacked in a matrix $X^p \in \mathbb{R}^{T \times p}$.
In a multi-scale setting, we will often consider multiple window sizes $\{p_j\}^J_{j=1}$ with
$p_j = 2^j$ depending on the case. To simplify notations we write $X_t^j$ to denote $X_t^{p_j}$.
This enables us to denote a family of geometrically increasing lag-windows with a parameter $j \in 1, 2, ...J$
where $J$ refers to the largest scale. An example time series with the corresponding lag-matrix is shown 
in figure~\ref{fig:lagmatrix example}.

\begin{figure*}
  \centering
  \includegraphics[width=0.9\textwidth]{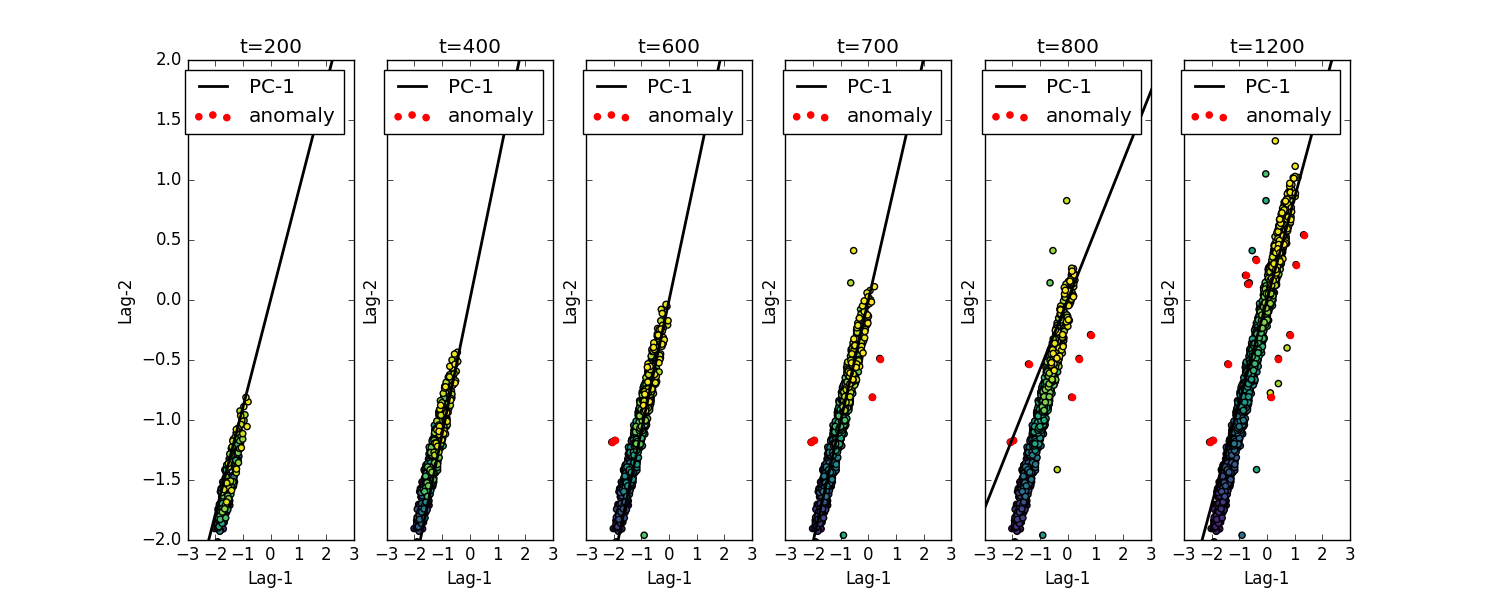}
  \includegraphics[width=0.6\textwidth]{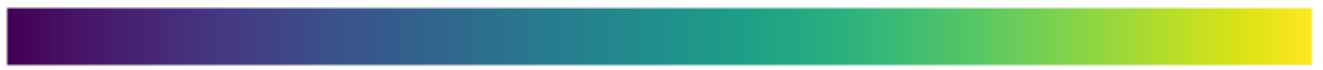}
  \caption{Evolution of the principal direction and the point cloud for a 2-dimensional lagmatrix.
  The color scale shown below represents (viewed from left to right) the direction of time. The black line
  in each plot represents the principal direction at that time instant $t$. One should also note that
  the direction is adapted even to anomalous points since they are never known in advance. This might lead
  to increased false positives.}
  \label{fig:pc_evolve}
\end{figure*}

We focus on a class of methods relying on the online estimation of the 
principal component of the lag-matrix, pioneered by~\cite{PapadimitriouSPIRIT2005}.
The anomaly score is then simply derived as the reconstruction error. For simplicity, 
we only consider the first principal direction.

Given a design matrix $X^p \in \mathbb{R}^{T \times p}$, the first principal 
direction $\mathbf{w}_p$ is defined as the 1-D projection capturing most of the energy of the data samples :

\begin{equation}
   \mathbf{w}_p = \arg\min_{\|\mathbf{w}\|=1} \sum^T_{t=1} 
   \| X_{t}^p - (\mathbf{w}\mathbf{w}^T)X_t^p \|^2
 \end{equation} 

We solve this problem in an online fashion following the approach of~\cite{PapadimitriouSPIRIT2005}.
For a given scale ~\cite{PapadimitriouSPIRIT2005} use the projective approximation subspace 
tracking algorithm (PAST) first introduced in ~\cite{Yang1995PastD}. 
For the formal proofs of convergence to the true principal subspace projection, 
we refer you to a good review provided by \cite{landqvist2005projection}.
An adapted version of this method can be read in Algorithm~\ref{algo:streaming_PCA}.

At time $t$, we denote by $\widetilde{X}_t^p$ the projection of $X_t^p$ 
upon $\mathbf{w}^p$ (at this time step), \textit{i.e.} $\widetilde{X}_t^p = \mathbf{w}_p^T X_t^p$.
The anomaly score at time $t$ and for a window of size $p$ will be written 
as $\alpha_t^p = \Vert \widetilde{X}_t^p - X_t^p \Vert^2$. When considering dyadic windows $p_j$, 
we also write $\mathbf{w}_{p_j} =: \mathbf{w}_j$ and $\alpha^{p_j}=:\alpha^j$.

\section{Streaming anomaly detection}

In this section we propose two algorithms for anomaly detection.
They develop on the Streaming PCA and Hierarchical PCA algorithms
developed for optimal pattern extraction in time series, and apply
them to the problem of anomaly detection, while being more amenable 
to multiple scales. We show an example of the application of the streaming
PCA algorithm in figure \ref{fig:pc_evolve} at scale=2. In this example
the lagmatrix has 2-dimensions that can be visualized in a plane. We see the
evolution of the principal direction with new incoming points. It is important
to note that for each new incoming point, we project it and its $p-1$ past samples, i.e. $X_t^p$
onto the current principal direction $\mathbf{w}_p$ and adapt the direction 
so as to reduce the error of reconstruction. This is not a prediction.
$\mathbf{w}_p \mathbf{w}_p X_t^p$ captures the largest amount of variance in the time series.
The number of principal directions determine the degrees of independent variation in the time series. 

\subsection{Multiscale Streaming PCA} 

The streaming PCA algorithm introduced by~\cite{PapadimitriouSPIRIT2005} consists in 
computing an online PCA on the past lags of the time series $X^j$. We use the subsequent 
reconstruction error as an anomaly score. Throughout this paper, we only consider the first 
PCA direction, though it is a straightforward extension to consider the first $k$ PCA directions.

A natural extension of this algorithm to a multiscale setting would consist in computing 
simultaneously the PCA directions at multiple scales $p_j = 2^j$ for $j \geq 1$ up to a 
maximal scale $P = 2^J$. The main drawback of this extension is its time complexity.
Indeed, for a past window of size $p_j$, the streaming PCA algorithm requires $\mathcal{O}(T p_j)$ 
operations to attribute an anomaly score on a time series of length $T$.
This multiscale extension would therefore require $\mathcal{O}(T P)$, which can be 
prohibitive for $P$ large. This is introduced in algorithm~\ref{algo:streaming_PCA}. 

\adjustbox{valign=t}{
\begin{minipage}{0.45\textwidth}
\begin{algorithm}[H]\small
\begin{algorithmic}
\State Initialization: $\mathbf{w}_j \leftarrow \mathbf{0}$, $\sigma_j^2 \leftarrow \epsilon$ with $\epsilon \ll 1$
\For{$t=1,\ldots, T$}
\For{$j=1, \ldots, J$}
\State $Z_t^j \leftarrow H_{2^j}^T X_t^j $
\State $y^j_t \leftarrow \mathbf{w}_j^T X_t^j$
\State $\sigma_j^2 \leftarrow \sigma_j^2 + (y_t^j)^2$
\State $\mathbf{e}_t^j \leftarrow Z_t^j - y_t^j \mathbf{w}_j$
\State $\mathbf{w}_j \leftarrow \mathbf{w}_j + \sigma_j^{-2} y_t^j \mathbf{e}_t^j$
\State $\pi_t^j \leftarrow \mathbf{w}_j^T Z_t^j$
\State $\widetilde{Z}_t^j \leftarrow \pi_t^j\mathbf{w}_j$
\State $\alpha_t^j \leftarrow \Vert \widetilde{Z}_t^j - Z_t^j \Vert^2$
\EndFor
\EndFor
\\
\Return $\pmb{\alpha} \in \mathbb{R}^{T \times J}$
\end{algorithmic}
\caption{Multiscale streaming PCA\label{algo:streaming_PCA}}
\end{algorithm}
\end{minipage}}

\subsection{Hierarchical streaming PCA}

\adjustbox{valign=t}{
\begin{minipage}{0.45\textwidth}
\begin{algorithm}[H]\small
\begin{algorithmic}
\State Initialization: $\mathbf{w}_j \leftarrow \mathbf{0}$, $\sigma_j^2 \leftarrow \epsilon$ with $\epsilon \ll 1$
\For{$t=1, \ldots, T$}
\For{$j=2, \ldots, J$}
\If{$j = 1$}
\State $Z_t^j \leftarrow H_{2^j}^T X_t^j $
\Else
\State $Z_t^j \leftarrow [\pi_t^{j-1}, (Z_t^{j})^T]$
\EndIf
\State $y_t^j \leftarrow \mathbf{w}_j^T Z_t^j$
\State $\sigma_j^2 \leftarrow \sigma_j^2 + (y_t^j)^2$
\State $\mathbf{e}_t^j \leftarrow Z_t^j - y_t^j \mathbf{w}_j$
\State $\mathbf{w}_j \leftarrow \mathbf{w}_j + \sigma_j^{-2} y_t^j \mathbf{e}_t^j$
\State $\pi_t^j \leftarrow \mathbf{w}_j^T Z_t^j$
\State $\widetilde{Z}_t^j \leftarrow \pi_t^j\mathbf{w}_j$
\State $\alpha_t^j \leftarrow \Vert \widetilde{Z}_t^j - Z_t^j \Vert^2$
\EndFor
\EndFor
\\
\Return $\pmb{\alpha} \in \mathbb{R}^{T \times J}$
\end{algorithmic}
\caption{Hierarchical streaming PCA\label{algo:hierarchical_PCA}}
\end{algorithm}
\end{minipage}
}

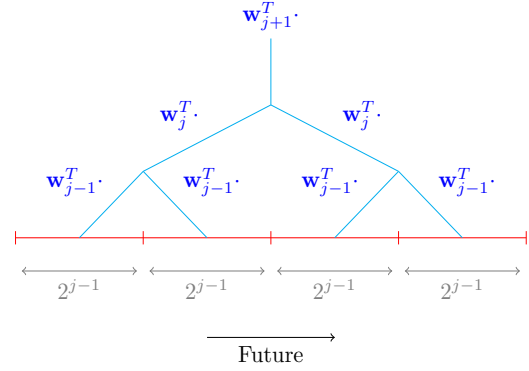
\begin{figure}[ht]
  \centering
  \resizebox{7cm}{5cm}{
  \begin{tikzpicture}

  \draw[red] (-9, 0) -- (-1, 0) ;
  \draw[red] (-9, 0.1) -- (-9, -0.1) ;
  \draw[red] (-7, 0.1) -- (-7, -0.1) ;
  \draw[red] (-5, 0.1) -- (-5, -0.1) ;
  \draw[red] (-3, 0.1) -- (-3, -0.1) ;
  \draw[red] (-1, 0.1) -- (-1, -0.1) ;
  
  \draw[<->, gray] (-8.9, -0.5) -- (-7.1, -0.5);
  \draw[gray] (-8, -0.5) node [below] {$2^{j-1}$};
  \draw[<->, gray] (-6.9, -0.5) -- (-5.1, -0.5);
  \draw[gray] (-6, -0.5) node [below] {$2^{j-1}$};
  \draw[<->, gray] (-4.9, -0.5) -- (-3.1, -0.5);
  \draw[gray] (-4, -0.5) node [below] {$2^{j-1}$};
  \draw[<->, gray] (-2.9, -0.5) -- (-1.1, -0.5);
  \draw[gray] (-2, -0.5) node [below] {$2^{j-1}$};
  
  \draw[->] (-6, -1.5) -- (-4, -1.5);
  \draw (-5, -1.5) node [below] {Future};
  
  \draw[cyan] (-8, 0) -- (-7, 1);
  \draw[blue] (-7.5, 0.5) node[above left] {$\mathbf{w}_{j-1}^T \cdot$};
  \draw[cyan] (-6, 0) -- (-7, 1); 
  \draw[blue] (-6.5, 0.5) node[above right] {$\mathbf{w}_{j-1}^T \cdot$};

  \draw[cyan] (-4, 0) -- (-3, 1);
  \draw[blue] (-3.5, 0.5) node[above left] {$\mathbf{w}_{j-1}^T \cdot$};
  \draw[cyan] (-2, 0) -- (-3, 1); 
  \draw[blue] (-2.5, 0.5) node[above right] {$\mathbf{w}_{j-1}^T \cdot$};
  
  \draw[cyan] (-7, 1) -- (-5, 2);
  \draw[blue] (-6, 1.5) node [above left] {$\mathbf{w}_j^T \cdot$};
  \draw[cyan] (-3, 1) -- (-5, 2);
  \draw[blue] (-4, 1.5) node [above right] {$\mathbf{w}_j^T \cdot$};  
  
  \draw[cyan] (-5, 2) -- (-5, 3);
  \draw[blue] (-5, 3) node [above] {$\mathbf{w}_{j+1}^T \cdot$};
  \end{tikzpicture}
  }
\caption{Hierarchical Streaming PCA.}
\label{fig:hierarchical_algos}
\end{figure}

To overcome redundant calculation in the multiscale streaming PCA algorithm 
\cite{papadimitriou2006optimal} introduced a hierarchical approximation, 
sketched in Figure~\ref{fig:hierarchical_algos} and algorithm~\ref{algo:hierarchical_PCA}.
When passing from one scale $p_j$ to the next $p_{j+1} = 2 p_j$, instead of 
rebuilding the lag matrix $X_t^{j+1}$ whose size doubles, it builds a reduced 
lag matrix $Z_t^{j+1}$ by considering the projection of each component of 
size $p_j$ on the principal direction obtained at this scale, \textit{i.e.}
$Z_t^{j+1} = [\mathbf{w}^T_j Z_t^j, \mathbf{w}^T_j Z_{t-2^j}^j]^T$ with $Z_t^1 = X_t^1$.
The principal direction at scale $p_{j+1}$ is then obtained by applying the streaming 
PCA algorithm on this reduced representation. Note that the reconstruction error at 
each scale is only relative to the previous scale, and we do not back-propagate to the
finest scale of the actual time series. In some sense, this approach can be seen as a 
linear convolutional neural network (CNN).

\begin{figure*}
  \centering
  \resizebox{14cm}{7cm}{%
  \begin{tikzpicture}[node distance=4cm,scale=0.50]
  \node (start) [startstop] {Input $x_t$};
  \node[text width=2cm,align=center] (in1) [io, right of=start] {Lagmatrix $X^{2^{j=1}}_t$};
  \node[text width=2cm,align=center] (in2) [io, below of=in1] {Lagmatrix $X^{2^{j=2}}_t$};
  \node[text width=2cm,align=center] (in3) [io, below of=in2] {Lagmatrix $X^{2^{j=J}}_t$};
  \node (dimred1) [preparation, right of =in1] {Haar Tx. $H_1$};
  \node (dimred2) [preparation, right of =in2] {Haar Tx. $H_2$};
  \node (dimred3) [preparation, right of =in3] {Haar Tx. $H_J$};  
  \node (pro1) [process, right of=dimred1] {Error $\alpha^{j=1}_t$};
  \node (pro2) [process, right of=dimred2] {Error $\alpha^{j=2}_t$};
  \node (pro3) [process, right of=dimred3] {Error $\alpha^{j=J}_t$};
  \node[text width=2cm,align=center] (dec1) [decision, right of=pro1] {Norm $\|\pmb{\alpha}_t\|^2$};
  \node[text width=2cm,align=center] (dec2) [decision, right of=pro2] {2nd iteration \\ $\|\widetilde{\pmb{\alpha}_t} - \pmb{\alpha}_t\|^2$};
  \node[text width=2cm,align=center] (dec3) [decision, right of=pro3] {Least correlated \\ scale $ j^\ast = {\arg\min}_j $\\$\sum_i (\pmb{\alpha}^T \pmb{\alpha})_{ji}$};

  \path (in2) -- node[auto=false]{\vdots} (in3);
  \path (dimred2) -- node[auto=false]{\vdots} (dimred3);
  \path (pro2) -- node[auto=false]{\vdots} (pro3);
  \path (dec2) -- node[auto=false]{\vdots} (dec3);
  
  \draw [arrow] (start) -- (in1);
  \draw [arrow] (start) -- (in2);
  \draw [arrow] (start) -- (in3);
  \draw [arrow] (in1) -- (dimred1);
  \draw [arrow] (in2) -- (dimred2);
  \draw [arrow] (in3) -- (dimred3);
  \draw [arrow] (dimred1) -- (pro1);
  \draw [arrow] (dimred2) -- (pro2);
  \draw [arrow] (dimred3) -- (pro3);
  \draw [dotted_arrow] (in1) -- (dimred2);
  \draw [dotted_arrow] (in2) -- (dimred3);
  \draw [arrow] (pro1) -- (dec1);
  \draw [arrow] (pro2) -- (dec1);
  \draw [arrow] (pro3) -- (dec1);
  \draw [arrow] (pro1) -- (dec2);
  \draw [arrow] (pro2) -- (dec2);
  \draw [arrow] (pro3) -- (dec2);
  \draw [arrow] (pro1) -- (dec3);
  \draw [arrow] (pro2) -- (dec3);
  \draw [arrow] (pro3) -- (dec3);
  \end{tikzpicture}
  }
  \caption{Schematic representation of the proposed method and different multi-scale 
  anomaly score aggregation methods. After the construction 
  of an adapted representation of the lag matrix at multiple scales, which can be 
  done in a hierarchical manner (indicated by the dotted arrows), anomaly scores are 
  obtained for each scale and then recombined to obtain a final score.}
  \label{fig:multiscale_scores_overview}
\end{figure*}
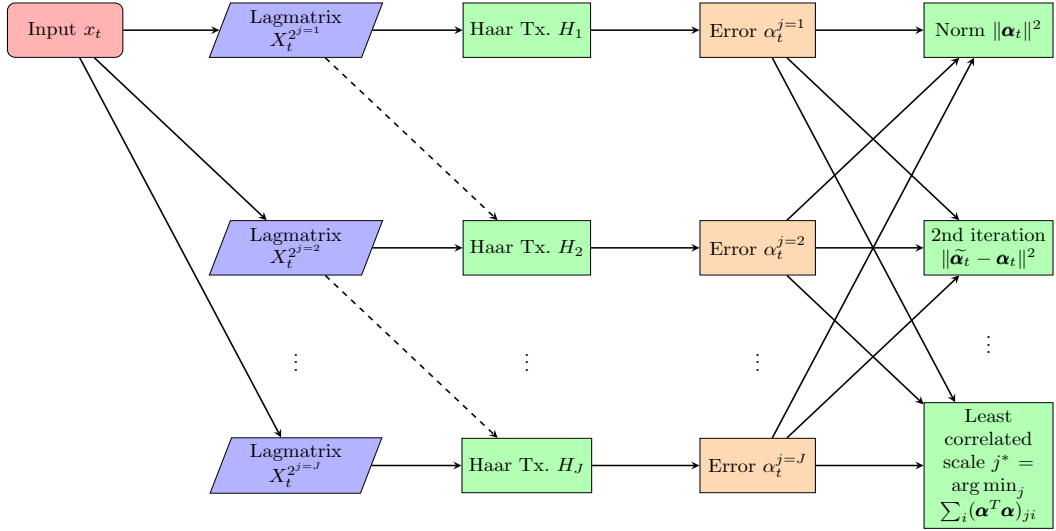
  
\begin{figure}
  \centering
  \includegraphics[width=0.45\textwidth]{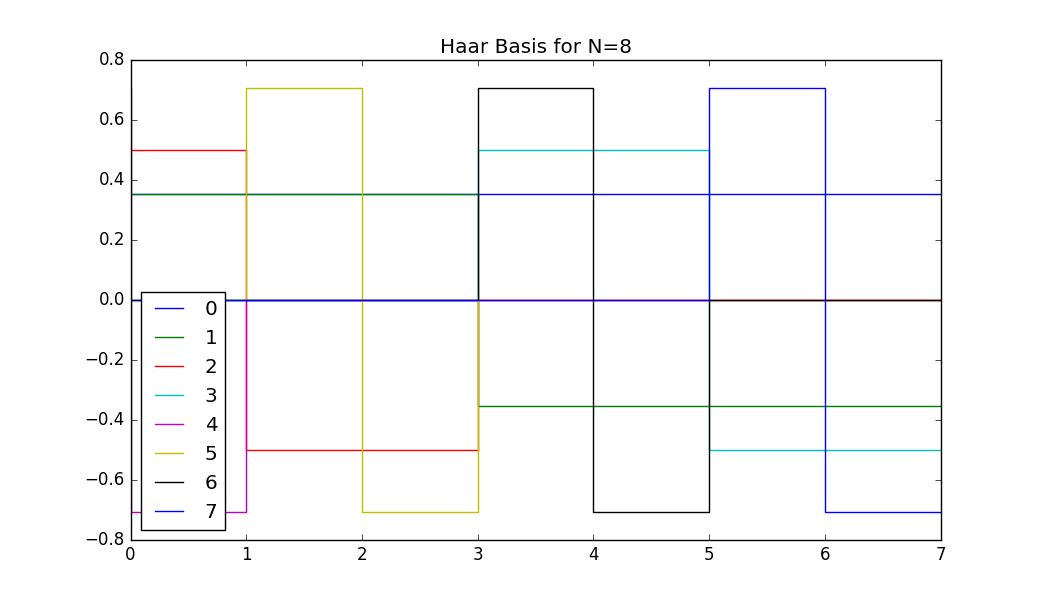}
  \caption{Haar basis for time series length N=8.}
  \label{fig:haar_example}
\end{figure}

\begin{figure*}
  \centering
  \includegraphics[width=0.275\textwidth]{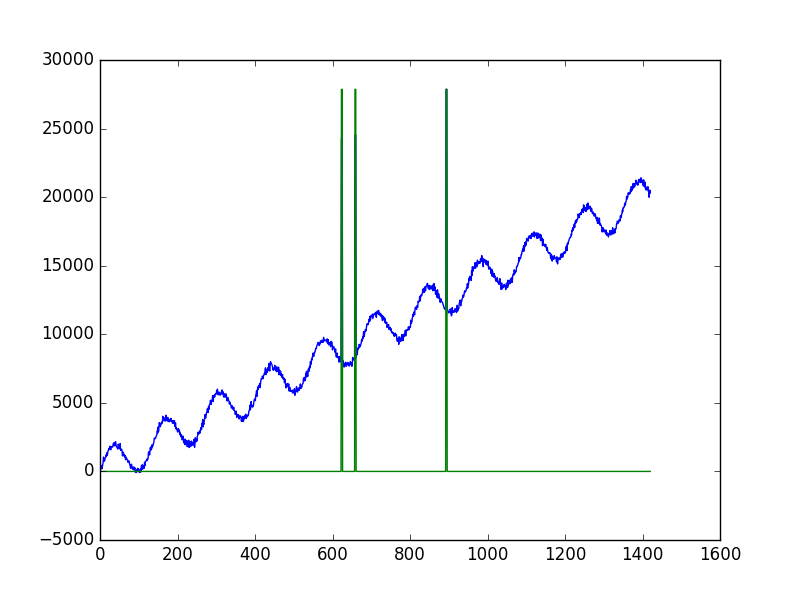}
  \includegraphics[width=0.7\textwidth]{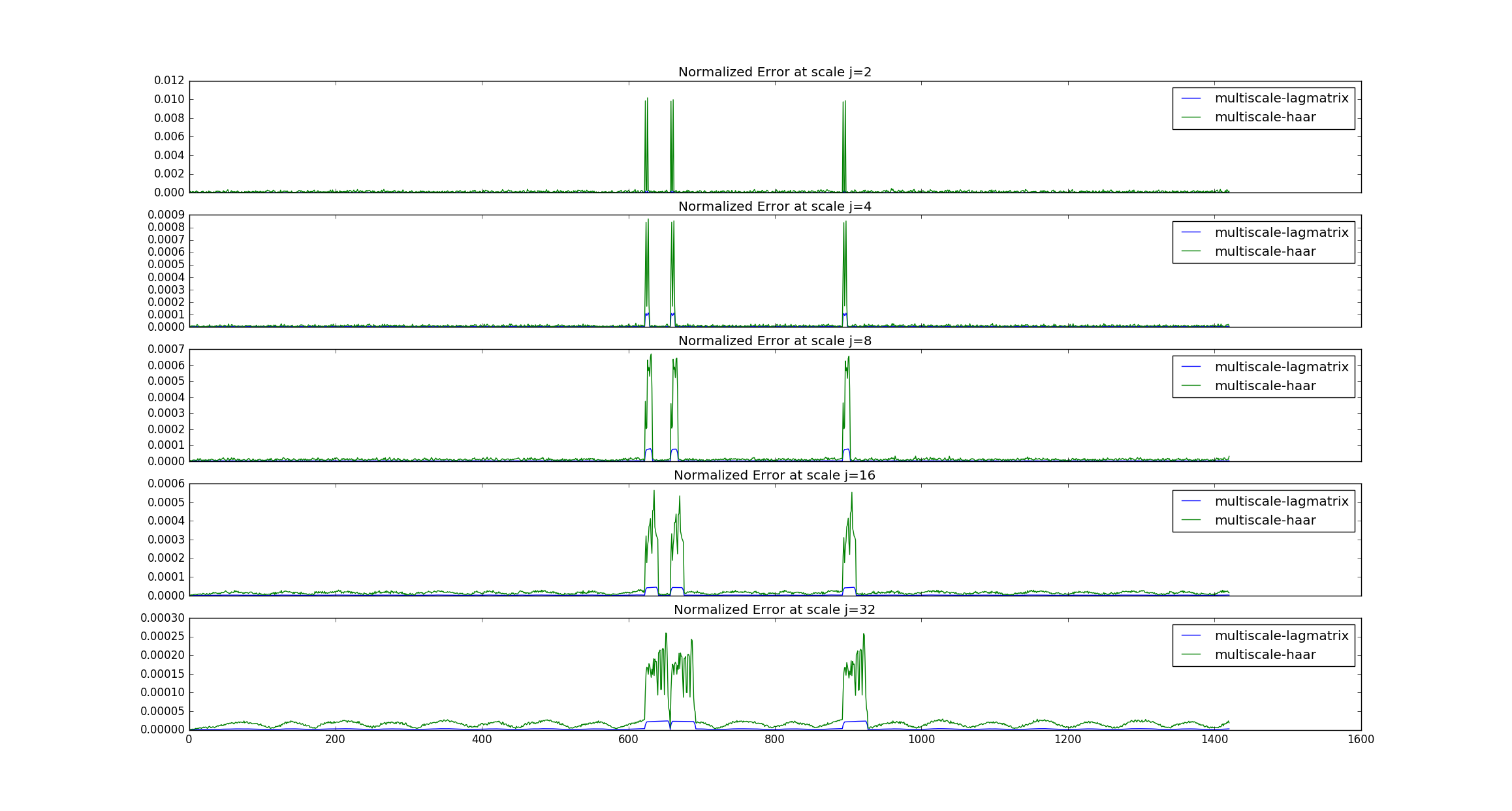}
  \caption{Example time series in blue with the anomalies market in green (left). 
  The errors of reconstruction (right) at different geometric scales.
  The reconstruction error for the change of basis with the haar transform is also plotted. The reconstruction
  error of the haar coefficients produces a shared peak in error compared in the reconstruction error of the
  naive lag-matrix.}
  \label{fig:recon_example}
\end{figure*}

The main assumption underlying this approach is that at each change of scale, 
the projection at the previous scale is equally valid for the near past (first 
half at scale $p_j$) and the far past (second half at scale $p_j$). In the 
context of streaming anomaly detection, we argue that this assumption is not 
necessarily relevant. Yet, the benefits of the hierarchy for run-time 
complexity should not be ignored: the Hierarchical PCA algorithm 
takes $\mathcal{O}(T \log P)$ operations to run up to scale $P$, where $P=2^J$.

\paragraph{Haar transform :} The haar transform is a multiscale 
approximation of an input time series window $X_t^p$
at of window size N=$2^p$ can be calculated by the dot product with the haar basis.
The unitary haar basis are calculated by the following recursion equation with $H_1 = [1]$ :

\begin{equation}
  H_{2N}={\frac{1}{\sqrt{2}}
  \begin{bmatrix}
  H_{N}\otimes [1,1]\\I_{N}\otimes [1,-1]
  \end{bmatrix}}
\end{equation}

The haar basis for a time series of length $N=4$ is given by : 
\begin{equation}
    H_{4}={\frac {1}{2}}
    {\begin{bmatrix}
    1 & 1 & 1 & 1\\
    1 & 1 & -1 & -1\\
    \sqrt{2} & -\sqrt{2} & 0 & 0\\
    0 & 0 & \sqrt{2} & -\sqrt{2}
    \end{bmatrix}}
\end{equation}

For each lag window $X_t^p$ we perform a change of basis using the unitary Haar transform $H^T X_t^p$.
This provides a representation of the time series that is well known in wavelet analysis
to approximate signals, the basis are shown in figure \ref{fig:haar_example}. 
In this study we simply use the complete change of basis to understand 
its effect on the reconstruction error. 
We observe in figure \ref{fig:recon_example}, an relatively sharper 
spike in error at the anomaly for a change of basis using
the Haar basis, as compared to a regular multiscale lagmatrix reconstruction.
In future work we aim to perform the Haar decomposition 
in streaming and perform a reconstruction of the time series, without evaluating the principal subspace. 

The final flow of operations per time sample is demonstrated in figure~\ref{fig:multiscale_scores_overview}.
The hierarchical approximation is not shown, but basically consists again in calculating the reconstruction
error of the multi-scale lagmatrix. The table~\ref{table:bigO_complexities} gives the time complexity for the 
streaming PCA algorithm as well as it's hierarchical approximation.

\begin{table}[h!]
\begin{center}
\begin{tabular}{|c|c|c|c|c| } 
 \hline
 Algorithm & Multiscale PCA & Hierarchical PCA \\
 \hline
 Complexity & $\mathcal{O}(T P)$ & $\mathcal{O}(T \log{} P)$  \\
 \hline
\end{tabular}
\end{center}
\caption{Number of floating point operations of streaming PCA for multiscale methods 
on a time series of length $T$ up to a maximal window size $P$. 
\label{table:bigO_complexities}}
\end{table}

\section{Experiments}
In this section we evaluate the performance of the different time series representations
for the streaming PCA based multiscale anomaly detector. In this light we also propose
three different ways of deciding the final score given an input anomaly score at 
multiple scales $\pmb{\alpha}_t = \lbrace \alpha_t^j \rbrace_{j \leq J}$. Namely, 
we evaluate the following aggregations:
\begin{enumerate} 
  \item $\| \pmb{\alpha}_t \|^2 $ : Norm of multiscale anomaly score
  \item $\| \widetilde{\pmb{\alpha}_t} - \pmb{\alpha}_t\|^2$: 
  Streaming reconstruction error on anomaly score, obtained \textit{via} a 2nd 
  iteration of the streaming PCA algorithm on the multiscale anomaly score 
  instead of the lag-matrix.
  \item $\alpha_t^{j^\ast} $ where $ j^\ast = {\arg\min}_j \sum_i (\pmb{\alpha}^T \pmb{\alpha})_{ji}$ : 
  the anomaly score corresponding to the scale which is least correlated with others.
\end{enumerate}

To assess the performance of a detector, we use the area under the receiver 
operators characteristics curve (AUC), which is calculated by integrating the 
curve of the False positive rate(FPR) \textit{vs} the True positive rate (TPR) 
obtained for all possible thresholds.
This AUC score is comprised between $0, 1$ where $0.5$ corresponds to the 
worst value, that is detection is equivalent to random guessing while $1$ or $0$ is a perfect detector
or perfect rejection. In the plot of
AUC for individual time series we plot the AUC=0.5 in dotted lines to remark the location of 
random guessing. 
We evaluate the AUC for each time series in each benchmark and provide the mean value
with its standard deviation for the AUC's of the time series in each benchmark.

The work in ~\cite{Laptev2015YahooBenchmark} evaluates off-line models and do not consider 
a streaming anomaly detection setup. While \cite{LavinA15_Numenta_NAB} is expressly created
for streaming anomaly detection time series, though we do not report an explicit comparison 
with their scores in this study which would consist of evaluating in the NAB platform.
But we do plan for a complete evaluation in the future. All code will be soon available online.

\textbf{Dataset description}: In our study we use the Yahoo! time series dataset 
introduced by \cite{Laptev2015YahooBenchmark}. It consists of four different 
datasets (Benchmark $1$ to $4$), each of them containing approximately $100$ time series. 
Benchmark 1 contains real time series data from Yahoo! traffic data, which are 
time series with a variety of different scales of repeating patterns, and is 
the most difficult benchmark to detect anomalies. Benchmarks 2 and 3 have synthetic 
time series which are mainly sinusoids with varying frequencies and varying levels 
of noise with relatively easier to detect anomalies. Benchmark 4 contains 
non-stationary artificial time series with oscillations and linear trends, containing 
not only anomalies but change-points. Change-points are not evaluated in this study.
Finally the Numenta anomaly detection dataset \cite{LavinA15_Numenta_NAB} like benchmark 4, contains
synthetic and real time series with hard-to-detect examples.

\paragraph{Observations}: The results of evaluation of the multiscale streaming PCA, 
and the hierarchical approximation is reported in table~\ref{table:AUC_results_AUC_finalscores}.
Each benchmark consists of 70-100 series, we report the median value of the AUC values across
the benchmark. We also report the deviation of the AUC scores across their median (Median absolute deviation).
Given the multiscale anomaly scores $\pmb{\alpha}_t$ we evaluate the performance of the three
ways to aggregate the multi-scale score : $\| \pmb{\alpha}_t \|^2, \|\widetilde{\pmb{\alpha}_t} - 
\pmb{\alpha}_t\|^2$ and $\alpha_t^{j^\ast}$, namely the norm, iterated streaming PCA
on the multiscale score $\|\pmb{\alpha}_t\|^2$ and finally the least correlated scale. 
We also evaluate the performance when tracking only the first principal subspace (PC=1), 
and that for the first two subspaces (PC=2), to observe the effect of preserving more variance.
Finally we tracked the principal subspace of a single scale naive lag-matrix at one scale, 
to compare it's performance w.r.t a multi-scale approach. 

In all benchmarks, the multi-scale anomaly scores perform better than their 
single scale counterpart. An important empirical result is the effect of de-correlation 
between scales. We see that w.r.t the norm of the multi-scale score the least correlated
scale performs better on almost all benchmarks. This effect is also reflected in the
usage of a 2nd-iteration of the streaming PCA on the multi-scale anomaly score, whose
reconstruction error norm represents the projections that are orthogonal to the principal 
direction across all scales.

Except for Benchmark $1$ where considering 
the norm of all anomaly scores appears to provide the best results, running another streaming 
PCA on top of these scores appears to be the best way to achieve good results.
We also observe that the increase in the number of principal components do not improve the performance
of the detector.

\begin{figure*}[htbp]
  \centering
  \includegraphics[width=0.45\textwidth]{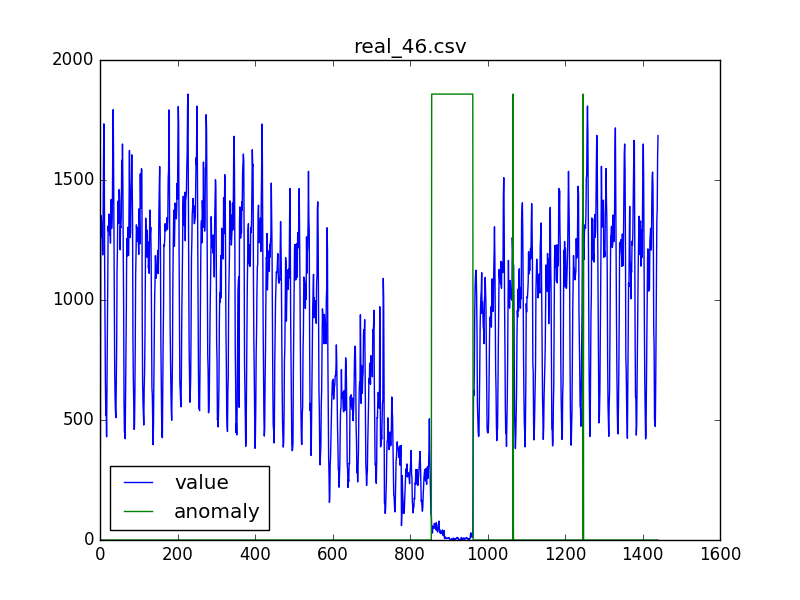}
  \includegraphics[width=0.45\textwidth]{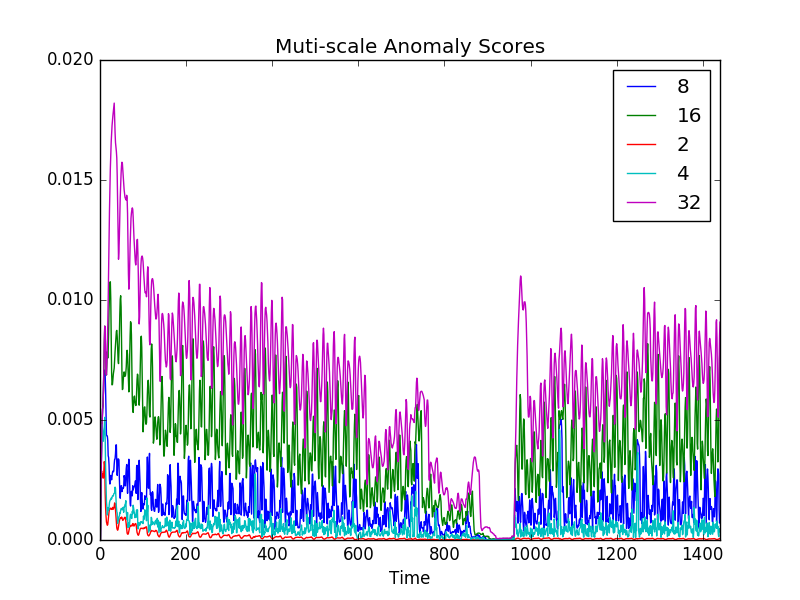}
  \caption{This example achieves AUC's of near 0 values in all three aggregation methods.
  We show the multiscale anomaly score. In such a case the reconstruction errors can be
  inverted, since the anomaly is a drop in reconstruction error.}
  \label{fig:failure}
\end{figure*}

\paragraph{Failure cases} In figure \ref{fig:failure} we have a time series
whose anomalous window is a one whose value drops. Reconstruction at different
scales are produce good approximations and no observable increase in errors, and thus
anomaly score. In such situations a non-linear function such as a minimum of 
moving window would have detected the anomaly.

\begin{figure*}
  \centering
  \includegraphics[width=0.95\textwidth]{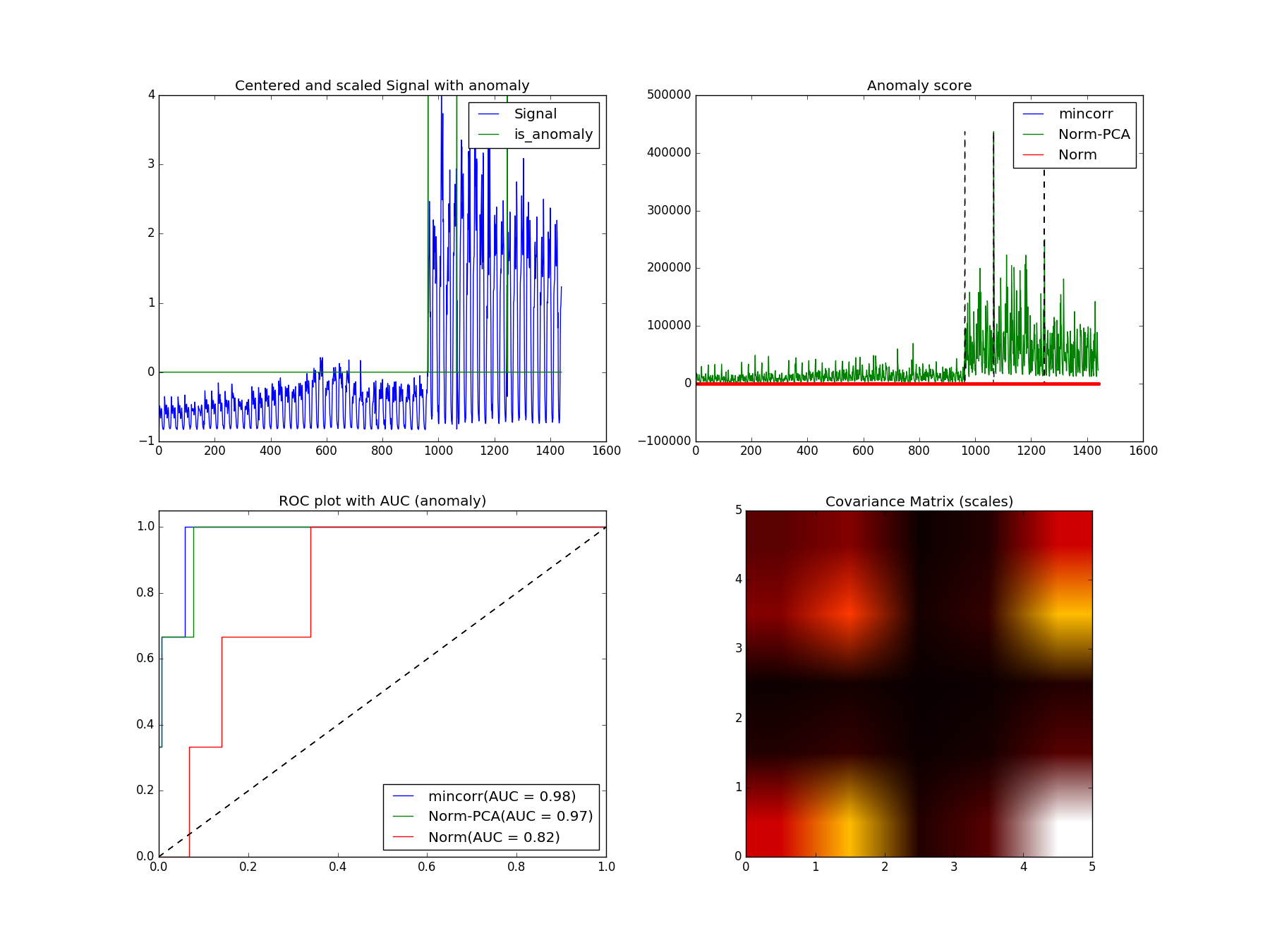}
  \caption{Time series with the three different final anomaly scores : Norm 
  $\| \pmb{\alpha}_t \|^2 $, Norm-PCA $\| \widetilde{\pmb{\alpha}_t} - \pmb{\alpha}_t\|^2$ (2nd iteration of PCA), 
  MinCorr / least correlated scale $\alpha_t^{j^\ast} $ where $ j^\ast = {\arg\min}~ \sum_i (\pmb{\alpha}^T \pmb{\alpha})_{ji}$. We also plot the covariance of the norm of the errors between different scales. The decorrelation
  in reconstruction errors across the different scales, in our study, produces a robust detection of anomalies.}
  \label{fig:all_scores_example}
\end{figure*}

\begin{table*}
    \label{table:AUC_results_AUC_finalscores}
    \begin{center} 
    \small
    \begin{tabular}{ |c|c|c|c|c|c| }
    
    \multicolumn{6}{c}{\colorbox{lightgray}{\bf Multi-scale score-Norm $\| \pmb{\alpha}_t \|^2 $ (PC=1)}} \\
    \hline
    \textbf{Method / AUCs}   & \textbf{Bench 1} & \textbf{Bench 2}  & \textbf{Bench 3}  & \textbf{Bench 4} & \textbf{NAB} \\ 
    \hline
    fixed-scale             &  $0.828\pm 0.240$ &  $0.835\pm 0.180$ &  $0.614\pm 0.108$ &  $0.568\pm 0.160$ &  $0.815\pm 0.238$ \\
    fixed-scale-haar        &  $0.826\pm 0.238$ &  $0.878\pm 0.143$ &  $0.617\pm 0.115$ &  $0.576\pm 0.157$ &  $0.812\pm 0.232$ \\
    multiscale-lagmatrix    &  $0.884\pm 0.232$ &  $0.978\pm 0.057$ &  $0.816\pm 0.092$ &  $0.696\pm 0.157$ &  $0.879\pm 0.199$ \\
    hierarchical-approx     &  $0.871\pm 0.236$ &  \colorbox{pink}{$0.997\pm 0.002$} &  $0.980\pm 0.025$ &  \colorbox{pink}{$0.897\pm 0.104$} &  \colorbox{pink}{$0.900\pm 0.189$} \\
    multiscale-haar         &  \colorbox{pink}{$0.906\pm 0.231$} &  $0.989\pm 0.019$ &  \colorbox{pink}{$0.992\pm 0.019$} &  $0.892\pm 0.126$ &  $0.892\pm 0.198$ \\
    \hline
    \multicolumn{6}{c}{\colorbox{lightgray}{\bf Multi-scale score-Norm $\| \pmb{\alpha}_t \|^2 $ (PC=2)}} \\
    \hline
    fixed-scale              &  $0.783\pm 0.269$ &  $0.918\pm 0.065$ &  $0.616\pm 0.142$ &  $0.569\pm 0.154$ &  $0.815\pm 0.231$ \\
    fixed-scale-haar         &  $0.808\pm 0.259$ &  $0.925\pm 0.074$ &  $0.627\pm 0.146$ &  $0.586\pm 0.144$ &  $0.811\pm 0.232$ \\
    multiscale-lagmatrix     &  $0.850\pm 0.242$ &  $0.969\pm 0.031$ &  $0.803\pm 0.116$ &  $0.686\pm 0.163$ &  $0.862\pm 0.210$ \\
    hierarchical-approx      &  $0.848\pm 0.240$ &  \colorbox{green}{$0.985\pm 0.056$} &  \colorbox{green}{$0.982\pm 0.021$} &  \colorbox{green}{$0.941\pm 0.079$} &  \colorbox{green}{$0.876\pm 0.213$} \\
    multiscale-haar          &  \colorbox{green}{$0.862\pm 0.245$} &  $0.976\pm 0.021$ &  $0.805\pm 0.150$ &  $0.710\pm 0.166$ &  $0.873\pm 0.195$ \\

    \hline
    \multicolumn{6}{c}{\colorbox{lightgray}{\bf PCA on multi-scale score $\| \widetilde{\pmb{\alpha}_t} - \pmb{\alpha}_t\|^2$ (PC=1)}}\\
    \hline
    \textbf{Method / AUCs}   & \textbf{Bench 1} & \textbf{Bench 2}  & \textbf{Bench 3}  & \textbf{Bench 4} & \textbf{NAB} \\ 
    \hline
    fixed-scale           &  $0.632\pm 0.264$ &  $0.754\pm 0.206$ &  $0.533\pm 0.124$ &  $0.525\pm 0.133$ &  $0.700\pm 0.247$ \\
    fixed-scale-haar      &  $0.649\pm 0.251$ &  $0.723\pm 0.194$ &  $0.514\pm 0.110$ &  $0.522\pm 0.129$ &  $0.699\pm 0.244$ \\
    multiscale-lagmatrix  &  \colorbox{pink}{$0.895\pm 0.218$} &  $0.997\pm 0.006$ &  \colorbox{pink}{$0.993\pm 0.017$} &  \colorbox{pink}{$0.959\pm 0.063$} &  \colorbox{pink}{$0.891\pm 0.194$} \\
    hierarchical-approx   &  $0.859\pm 0.233$ &  \colorbox{pink}{$0.997\pm 0.002$} &  $0.961\pm 0.071$ &  $0.895\pm 0.108$ &  $0.884\pm 0.204$ \\
    multiscale-haar       &  $0.888\pm 0.219$ &  $0.988\pm 0.031$ &  $0.956\pm 0.059$ &  $0.898\pm 0.106$ &  $0.886\pm 0.178$ \\
    \hline
    \multicolumn{6}{c}{\colorbox{lightgray}{\bf PCA on multi-scale score $\| \widetilde{\pmb{\alpha}_t} - \pmb{\alpha}_t\|^2$ (PC=2)}} \\
    \hline
    fixed-scale           &  $0.778\pm 0.270$ &  $0.908\pm 0.091$ &  $0.609\pm 0.133$ &  $0.573\pm 0.154$ &  $0.813\pm 0.232$ \\
    fixed-scale-haar      &  $0.804\pm 0.261$ &  $0.922\pm 0.079$ &  $0.625\pm 0.148$ &  $0.584\pm 0.143$ &  $0.811\pm 0.232$ \\
    multiscale-lagmatrix  &  $0.828\pm 0.237$ &  $0.872\pm 0.134$ &  $0.834\pm 0.172$ &  $0.793\pm 0.181$ &  $0.829\pm 0.207$ \\
    hierarchical-approx   &  \colorbox{green}{$0.831\pm 0.248$} &  \colorbox{green}{$0.978\pm 0.084$} &  \colorbox{green}{$0.976\pm 0.031$} &  \colorbox{green}{$0.935\pm 0.084$} &  \colorbox{green}{$0.841\pm 0.231$} \\
    multiscale-haar       &  $0.816\pm 0.239$ &  $0.933\pm 0.088$ &  $0.859\pm 0.161$ &  $0.799\pm 0.171$ &  $0.807\pm 0.226$ \\
    \hline
    \multicolumn{6}{c}{\colorbox{lightgray}{\bf Least correlated scale $\alpha_t^{j^\ast} $ where $ j^\ast = {\arg\min}_j \sum_i (\pmb{\alpha}^T \pmb{\alpha})_{ji}$ (PC=1)}}\\
    \hline
    \textbf{Method / AUCs}   & \textbf{Bench 1} & \textbf{Bench 2}  & \textbf{Bench 3}  & \textbf{Bench 4} & \textbf{NAB} \\ 
    \hline
    fixed-scale            &  $0.828\pm 0.240$ &  $0.835\pm 0.180$ &  $0.614\pm 0.108$ &  $0.568\pm 0.160$ &  $0.815\pm 0.238$ \\
    fixed-scale-haar       &  $0.826\pm 0.238$ &  $0.878\pm 0.143$ &  $0.617\pm 0.115$ &  $0.576\pm 0.157$ &  $0.812\pm 0.232$ \\
    multiscale-lagmatrix   &  $0.816\pm 0.238$ &  $0.773\pm 0.236$ &  \colorbox{pink}{$0.993\pm 0.017$} &  \colorbox{pink}{$0.964\pm 0.055$} &  $0.885\pm 0.196$ \\
    hierarchical-approx    &  $0.816\pm 0.238$ &  $0.773\pm 0.236$ &  $0.993\pm 0.017$ &  $0.964\pm 0.055$ &  $0.885\pm 0.196$ \\
    multiscale-haar        &  \colorbox{pink}{$0.832\pm 0.238$} &  \colorbox{pink}{$0.997\pm 0.007$} &  $0.799\pm 0.120$ &  $0.817\pm 0.123$ &  \colorbox{pink}{$0.886\pm 0.183$} \\
    \hline
    \multicolumn{6}{c}{\colorbox{lightgray}{\bf Least correlated scale $\alpha_t^{j^\ast} $ where $ j^\ast = {\arg\min}_j \sum_i (\pmb{\alpha}^T \pmb{\alpha})_{ji}$ (PC=2)}}\\
    \hline
    fixed-scale            &  $0.783\pm 0.269$ &  $0.918\pm 0.065$ &  $0.616\pm 0.142$ &  $0.569\pm 0.154$ &  \colorbox{green}{$0.815\pm 0.231$} \\
    fixed-scale-haar       &  \colorbox{green}{$0.808\pm 0.259$} &  \colorbox{green}{$0.925\pm 0.074$} &  \colorbox{green}{$0.627\pm 0.146$} &  $0.586\pm 0.144$ &  $0.811\pm 0.232$ \\
    multiscale-lagmatrix   &  $0.685\pm 0.332$ &  $0.757\pm 0.225$ &  $0.555\pm 0.140$ &  \colorbox{green}{$0.597\pm 0.168$} &  $0.736\pm 0.327$ \\
    hierarchical-approx    &  $0.689\pm 0.333$ &  $0.757\pm 0.225$ &  $0.555\pm 0.140$ &  $0.596\pm 0.167$ &  $0.736\pm 0.327$ \\
    multiscale-haar        &  $0.739\pm 0.318$ &  $0.765\pm 0.241$ &  $0.533\pm 0.200$ &  $0.512\pm 0.200$ &  $0.736\pm 0.336$ \\
   \hline
  \end{tabular}
  \end{center}
 \caption {AUC (mean $\pm$ standard deviation) for each benchmark, multiscale algorithm and aggregation method.
 We see that compared a fixed scale method a multi-scale method performs better. The effect of 
 the change of basis with haar improves the peakiness and thus the detection of anomalies, though
 this is still to be understood mathematically. The least correlated scale performs better than a 2nd iteration 
 of streaming PCA, which performs better on average compared to the norm of the multi-scale anomaly score.
 The decorrelation between scales represents an essential portion of anomalies in the datasets. 
 The (PC=2) refers to the AUC scores for the detection when two principal components are used to 
 reconstruct the lag-window. We observe that there is no increase in performance from PC=1(pink) to PC=2(green).
 }
\end{table*}

\section{Conclusion}

In this preliminary empirical study we observe the effect of scale on the reconstruction 
error in the streaming PCA set up for anomaly detection. We studied the effect of using
fixed window size to evaluate the reconstruction error by learning the first principal component, 
versus its multiscale counter-part. We observe that the performance of the detection
is also dependent on the different aggregation scores that are used.

\paragraph{Future work} : 
\begin{itemize}
  \item Firstly, we would like to understand the bounds on 
the reconstruction errors using the streaming PCA algorithm \cite{karnin2015online}, \cite{boutsidis2015online}.
The online PCA algorithm's target dimension and error bound are related in this study. 
\item Secondly we would like to understand the effect of temporal scale in higher dimensional time series, 
as well as the interaction between the scales of different components of a vectorial time series. 
The current paper has only studied the embedding of univariate time series viewed as a multivariate time series.
\item Thirdly it would useful to understand the effect of the representation of the
lag-window $W^T X_t^j$ in the hierarchical approximation of the multiscale PCA 
on the reconstruction error bound. 
\item Fourthly the lag parameter $p$ only represents 
the number of past/lag values of time series to be considered, though another degree
of liberty in time series is the embedding dimension (well know in the dynamical systems context) 
$\tau$ which refers to the number of samples to be skipped to get the next time window.
With increase in $\tau$ the windows become increasingly decorrelated.
\item Fifthly, the haar wavelet representation of the time series can be recursively
calculated for longer past-windows. This would enable us to model change in 
mean values w.r.t the far past.
\item Finally, in figure \ref{fig:pc_evolve} we see that the anomalous points 
have a significant effect on the change of the principal direction $\mathbf{w}_p$.
It would of interest to study a way to stop the adaptation of the principal direction 
when the reconstruction error is beyond a threshold, while still maintaining a high
anomaly score. This would provide a robust estimate to $\mathbf{w}_p$ while still detecting
the anomaly.
\end{itemize}

\paragraph{Acknowledgements} : The author would like to thank Prof. Stephane Mallat, ENS Paris, for 
discussions and supervision on this project, as well as Mathieu Andreux, ENS Paris, for all his help, 
comments and feedback on the project. I would like to acknowledge the support of Marc Tommasi 
and MAGNET-INRIA, in publishing this work.

I would also like to thank Rachid Ouaret from UPEC, and Senthil Yogamani from Valeo Ireland,
for their comments and feedback on the article.
\bibliography{cap2017_bib}

\end{document}